\documentclass[fleqn,usenatbib]{mnras}

\usepackage[T1]{fontenc}
\usepackage{ae,aecompl}
\usepackage{comment}
\usepackage{csquotes}
\usepackage[dvipsnames]{xcolor}

\DeclareRobustCommand{\VAN}[3]{#2}
\let\VANthebibliography\thebibliography
\def\thebibliography{\DeclareRobustCommand{\VAN}[3]{##3}\VANthebibliography}

\usepackage{mathtools}
\usepackage[varg]{txfonts}
\usepackage{color}
\usepackage{natbib}
\usepackage{multirow}   
\usepackage{gensymb}    
\usepackage{graphicx}
\graphicspath{ {Figures/} }
\usepackage{url}
\usepackage{indentfirst}
\usepackage[normalem]{ulem}     
\usepackage{placeins}
\usepackage{tablefootnote}
\usepackage{amsmath}	
\usepackage{amssymb}	
\hypersetup{colorlinks=true,allcolors=blue}
\newcommand{\bh}{H~1743-322 }
\newcommand{\nth}{\textit{Nthcomp }}
\newcommand{\hr}{$\dot{H}$ }
\newcommand{\tin}{$kT_{in}$}
\defcitealias{akash_garg2020identifying}{GA20}
\defcitealias{garg2022}{GA22}

\title[Understanding the variability of H~1743-322]{Investigating the Energy-Dependent Temporal Nature of Black Hole Binary System \bh}

\author[Husain et al. 2023]{
Nazma Husain,$^{1}$
Akash Garg,$^{1,2}$
Ranjeev Misra,$^{2}$
Somasri Sen$^{1}$
\\
$^{1}$Department of Physics, Jamia Millia Islamia, New Delhi-110025, India\\
$^{2}$Inter-University Centre for Astronomy and Astrophysics, Pune-411007, India\\
}

\date{Accepted XXX. Received YYY; in original form ZZZ}

\pubyear{2021}
\begin{document}
\label{firstpage}
\maketitle

\begin{abstract}


Black hole X-ray binaries routinely exhibit Quasi Periodic Oscillations (QPOs) in their Power density spectrum. Studies of QPOs have demonstrated immense ability to understand these dynamical systems although their unambiguous origin still remains a challenge. We investigate the energy-dependent properties of the Type-C QPOs detected for \bh as observed with \textit{AstroSat} in its two X-ray outbursts of 2016 and 2017. The combined broadband \textit{LAXPC} and \textit{SXT} spectrum is well modelled with a soft thermal and a hard Comptonization component. The QPO exhibits soft/negative lags i.e. variation in soft band lags the variation in hard band, although the upper harmonic shows opposite behaviour i.e. hard/positive lags. Here, we model energy-dependent properties (fractional root mean square and time-lag variation with energy) of the QPO and its upper harmonic individually with a general scheme that fits these properties by utilizing the spectral information and consequently allows to identify the radiative component responsible for producing the variability. Considering the truncated disk picture of accretion flow, a simple model with variation in inner disk temperature, heating rate and fractional scattering with time delays is able to describe the fractional RMS and time-lag spectra. In this work, we show that this technique can successfully describe the energy-dependent features and identify the spectral parameters generating the variability.

\vspace{0.3cm}
\end{abstract}


\section{Introduction}

The transient X-ray source H~1743-322, which is also a microquasar, was first detected during its 1977-78 outburst by all sky monitor Ariel V \citep{detect_kaluzienski1977iau} and HEAO I satellite \citep{detect_doxsey1977h}. Spectral and timing properties of this system such as its soft thermal emission (1-10~keV), high energy powerlaw tail (up to $\sim$100~keV), detection of high frequency variability and its spectral state transitions during an outburst indicate that the system harbours a black hole \citep{grade_A_mcclintock2003black}. Its black hole mass is still not dynamically confirmed although based on a mass scaling method given by \cite{grade_A_mcclintock2003black},  \cite{mass_mondal2009black} constrained it to be in range 6.6–11.3~$M_{\odot}$. \cite{steiner2012} estimated its distance to be $\sim$10~kpc by modelling the trajectories of jet and found it to be a  high inclination source with i=$75\degree$ $\pm$ $3\degree$, which is also supported by detection of strong dips in the lightcurve \citep{dips_homan2005high} similar to those found for other high inclination sources like GRO J1655-40 and 4U 1630-47. Moreover, \cite{steiner2012} constrained the black hole spin to be $a^{*}$ $\sim$0.2-0.3 by continuum-fitting method. Since its detection, the source has gone through several outbursts both full i.e. following the q-shaped trajectory in the Hardness-Intensity Diagram and failed i.e. partially covering the q-track \citep{partial_q_shape_sonbas2022temporal}. The spectral states are classified based on differing contributions coming from the accretion disk and the hot inner flow (medium of high energetic electrons), for classification of states see e.g. \cite{states_belloni2011black}. This binary system also exhibits fast X-ray variability such as the band limited noise and Quasi Periodic Oscillations (QPOs) which appear as narrow peaks in the power density spectrum (PDS). QPOs are considered to be produced in close vicinity of compact object and thus can reveal crucial information about matter falling into the strong gravitational force regime. Low-Frequency QPOs are generally classified into three types : A, B and C \citep{casella2005abc}; our interest here lies in Type-C QPOs which are most commonly observed and are also the strongest with rms of upto 15$\%$. These have been detected at low frequencies of $\sim$0.01~Hz and at more than 10~Hz for this system, sometimes accompanied with a harmonic as well as a sub-harmonic component. Type-C QPOs are usually detected during the hard and intermediate states of the outburst \citep{qpo_detect_molla2017}.

Even after decades of studying QPOs, there is still no consensus on how these (and in particular the Type-C QPOs which are the focus of this work) are produced in the inner region of the accretion flow. Till date, several models have been proposed to understand their origin, among which many consider a geometric origin for the QPOs where the oscillating geometry introduces quasi-periodic variability in the flux. Relativistic Precession Model (RPM) by \cite{origin_stella1997lense,origin_stella1999correlations} gives such a geometric origin by relating the QPO frequency to the Lense-Thirring Precession frequency at a characteristic radius, caused due to the frame-dragging effect of the spinning black hole. In model of \cite{origin_ingram2009low} the entire hot-inner flow precesses as a solid body in the framework of truncated disk geometry \citep{done2007} and the QPO is linked to this precessional frequency. For more details see \cite{ingram2011}. On the other hand, there are models which consider an intrinsic origin of QPOs in which the flux intrinsically varies quasi-periodically due to instabilities generated in the flow. For example,  \cite{origin_tagger1999accretion} presents a model where the magnetohydrodynamic instability in the disk causes spiral waves to travel and form standing patterns, which could be associated with these QPOs. Similarly, \cite{origin_titarchuk2000global} show that global disk oscillations due to the gravitational interaction of the compact object with the disk can also be the origin of QPOs. \cite{misra2020} have shown that the QPO could be identified as the dynamical frequency (defined as $f_{Dyn}$ = c(s)/r with c(s) being the sound speed) at the truncation radius of the disk (for more models on intrinsic origin see \cite{origin_chakrabarti2008evolution,cabanac2010variability}).

The underlying dynamical process which picks up a characteristic frequency of the flow component and turns it into a QPO still remains unresolved. QPO properties have been found to evolve as the source goes through the cycle of an outburst transitioning into different states, it points towards a seeming connection of the radiative processes with the QPO production. On that account, it is crucial to find out what role do the spectral parameters play in generating this distinctive feature. To accomplish this, a general approach has been taken into consideration in various works (these are discussed shortly) where small amplitude variation in spectral parameters is used to explain the energy-dependent properties of the QPO such as the variation of fractional amplitude and phase lags with energy. This approach considers an intrinsic origin for the QPOs. \cite{misra2013} present such a general framework where a driving oscillation generates variation in at least two dependent spectral parameters which vary with a phase lag changing the spectral flux. Using this technique, they further predict that assuming the spectrum to be completely described with an exponentially cut-off powerlaw the alternating nature of phase lags corresponding to one fundamental and its three harmonics of 65~mHz QPO for GRS 1915+105 could be described with variation produced in the powerlaw index and the inverse of cut-off energy. \cite{mir2016model} also utilized this approach and considered a delayed response of inner disk radius (DROID) to the sinusoidal oscillation generated in the accretion rate and applied it to QPO variability of GRS 1915+105. They show that the fractional RMS and time-lag variations can be completely described by only four parameters pertaining to the oscillation.

On similar line of work, \cite{akash_garg2020identifying} (Hereafter, \citetalias{akash_garg2020identifying}) formulated a model to describe the energy-dependent properties of QPO by variation of physical parameters of the accretion flow corresponding to disk blackbody and thermal comptonization emission, with the independence of choosing variation in any physical parameter and their corresponding time delays. They also show the applicability of their model to fit the temporal behaviour of GRS 1915+105. In their recent work, \cite{garg2022} (Hereafter, \citetalias{garg2022}) employed the same technique to the QPO observed for MAXI J1535-571. They find out that variation of mass accretion rate, inner disk radius and heating rate could well fit the fractional Root Mean Square (fRMS from hereon) and time-lag spectra.  Here, we make use of their model and try to identify radiative components required for generation of energy-dependent variability of QPO and its harmonic. This method requires knowledge of both spectral and timing properties of the source. Since most of the QPOs have been detected with \textit{RXTE} whose broadband spectrum range is limited to 2-20~keV, in this way, \textit{AstroSat} provides a broader range of 0.7-7.0~keV (\textit{SXT}) and 3.0-80.0~keV (\textit{LAXPC}).  This allows to account better for the disk emission which peaks at low energies $<$2~keV, also the timing resolution of \textit{LAXPC} is ideal to study properties at these frequencies.  Therefore, in this work, we analyse \bh with \textit{AstroSat} data and employ the technique described in \citetalias{akash_garg2020identifying,garg2022} to fit the energy-dependent properties of the QPO and its harmonic. 

The flow of the paper is as follows, in next Section we lay out details of observations with \textit{AstroSat} and extraction of data products.  In Section 3 we describe the spectral modelling and techniques used to study timing properties. We also explain the model and the fitting of fRMS and time-lag variation. Lastly, in Section 4 we discuss and interpret the obtained results.

\begin{figure}
    \hspace{-0.5cm}
    \includegraphics[width=10cm,height=5.5cm]{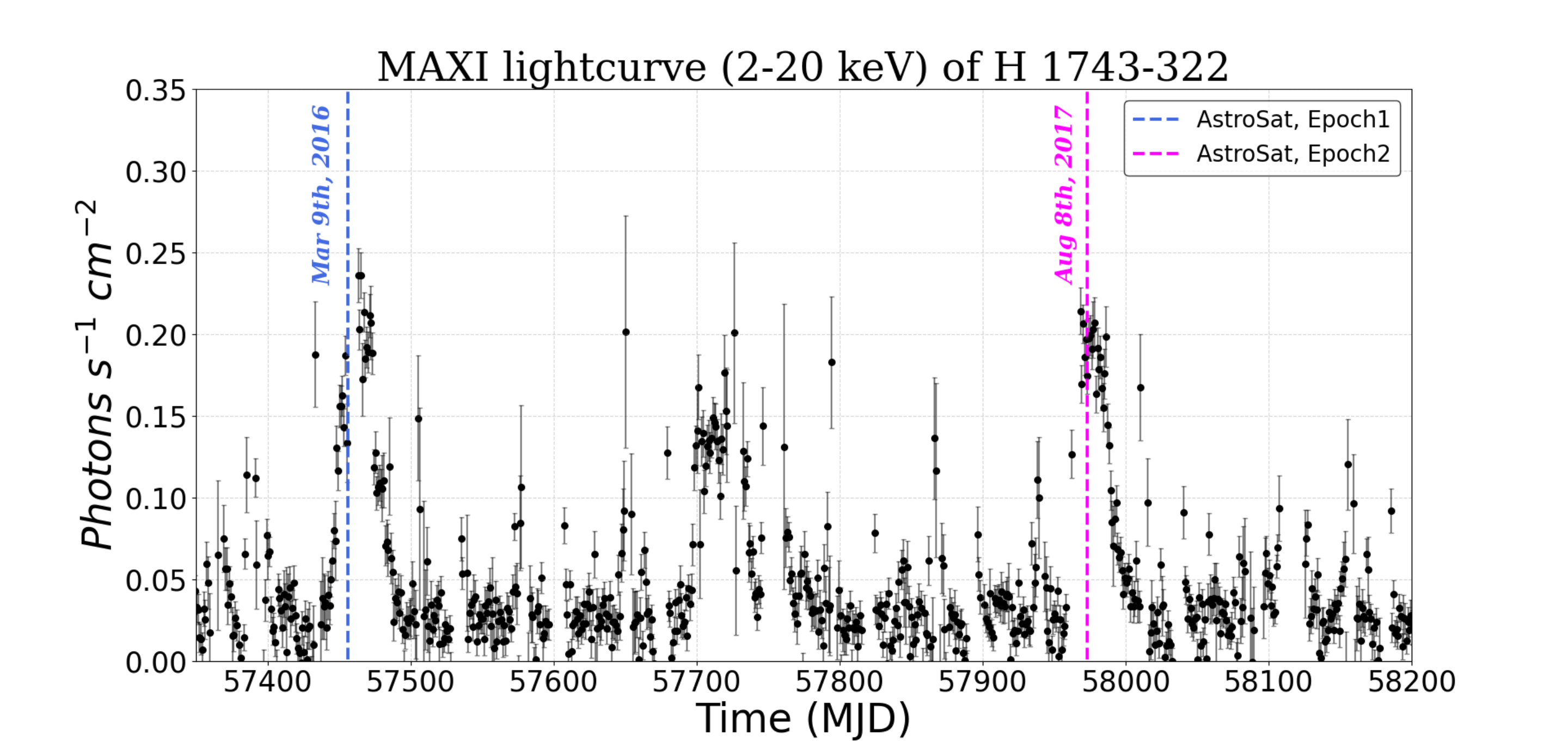}
    \caption{\textit{MAXI} lightcurve of \bh in 2-20~keV energy band. The vertical lines represent simultaneous \textit{AstroSat} observations with Epoch1 (blue) and Epoch2 (magenta). Both observations were conducted close to the peak  of each outburst.}
    \label{fig:maxi}
\end{figure}

\begin{figure*}
    \includegraphics[height=5.5cm, width=0.47\linewidth]{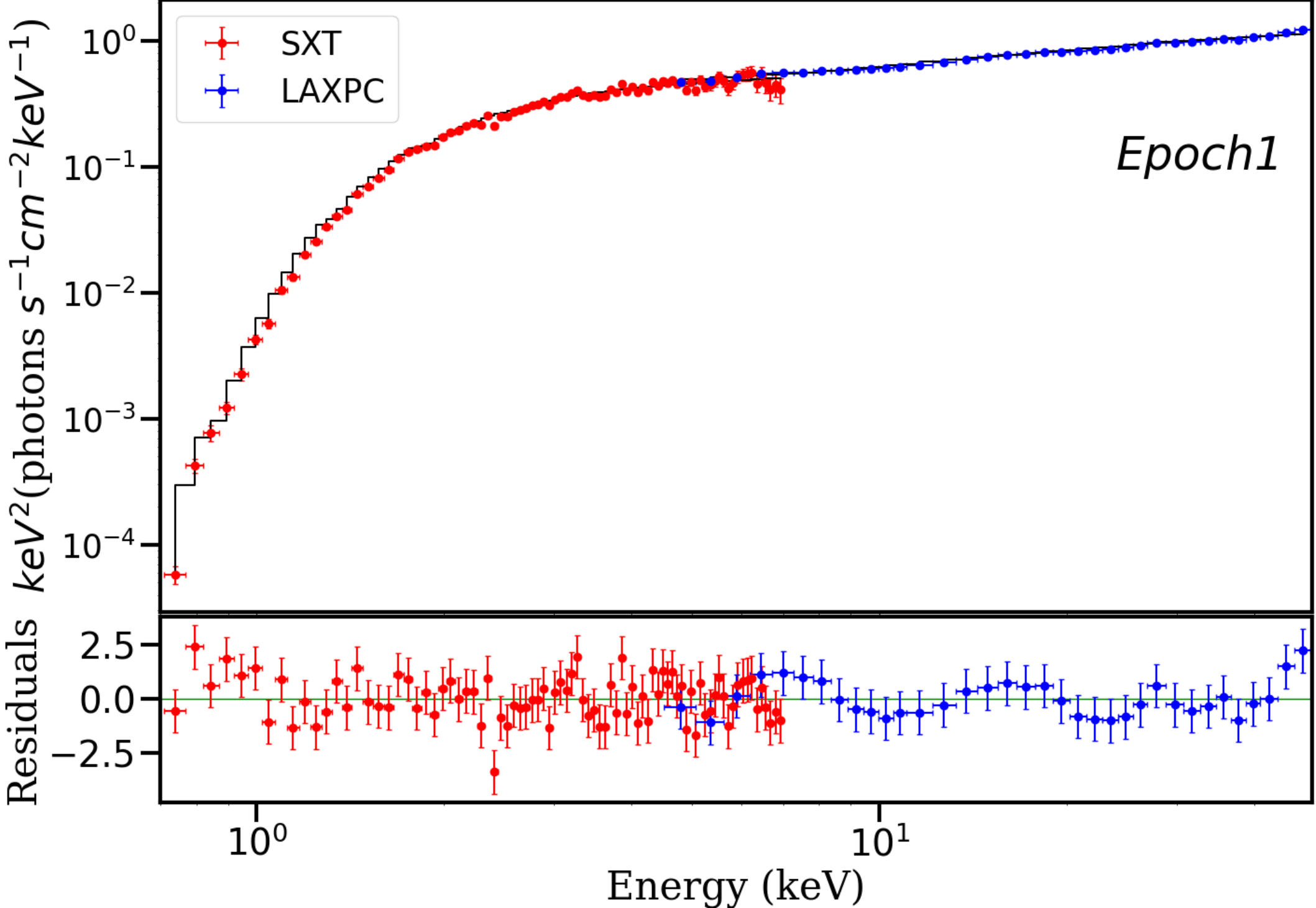}
    \hspace{0.5cm}
     \includegraphics[height=5.5cm, width=0.47\linewidth]{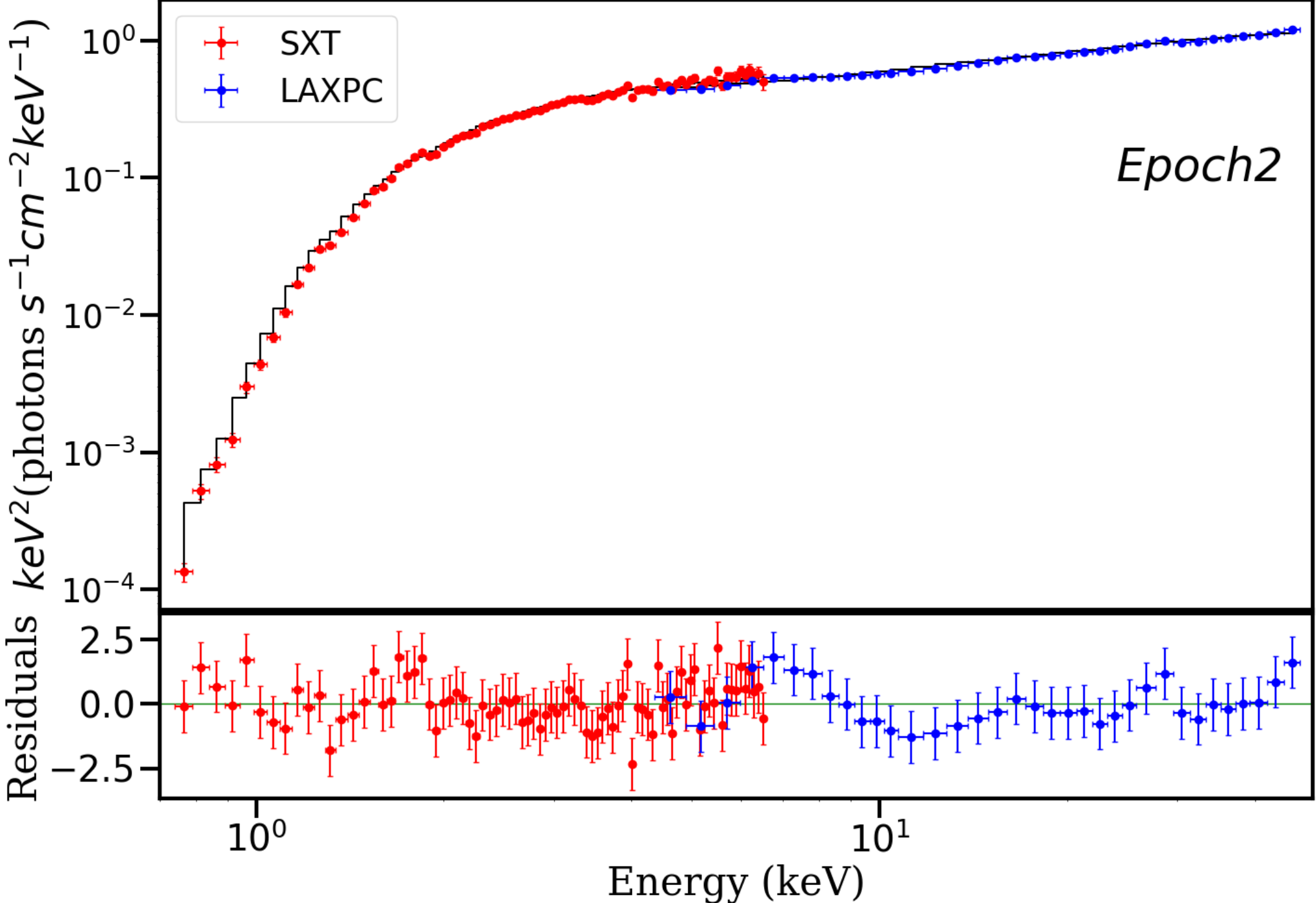}\\
    \caption{Broadband spectra of LAXPC (denoted with blue) and SXT (denoted with red) in energy range of 0.7-50~keV shown for Epoch1 (Left Panel) and Epoch2 (Right Panel). The joint spectra is fitted with model TBabs$\times$(Thcomp$\otimes$Diskbb) and the residuals are shown in the corresponding lower panel.}
    \label{fig:spectral_fit}
\end{figure*}

\begin{figure*}
    \includegraphics[trim=0cm 0cm 2cm 0cm, clip=true, height=6.5cm, width=0.49\linewidth]{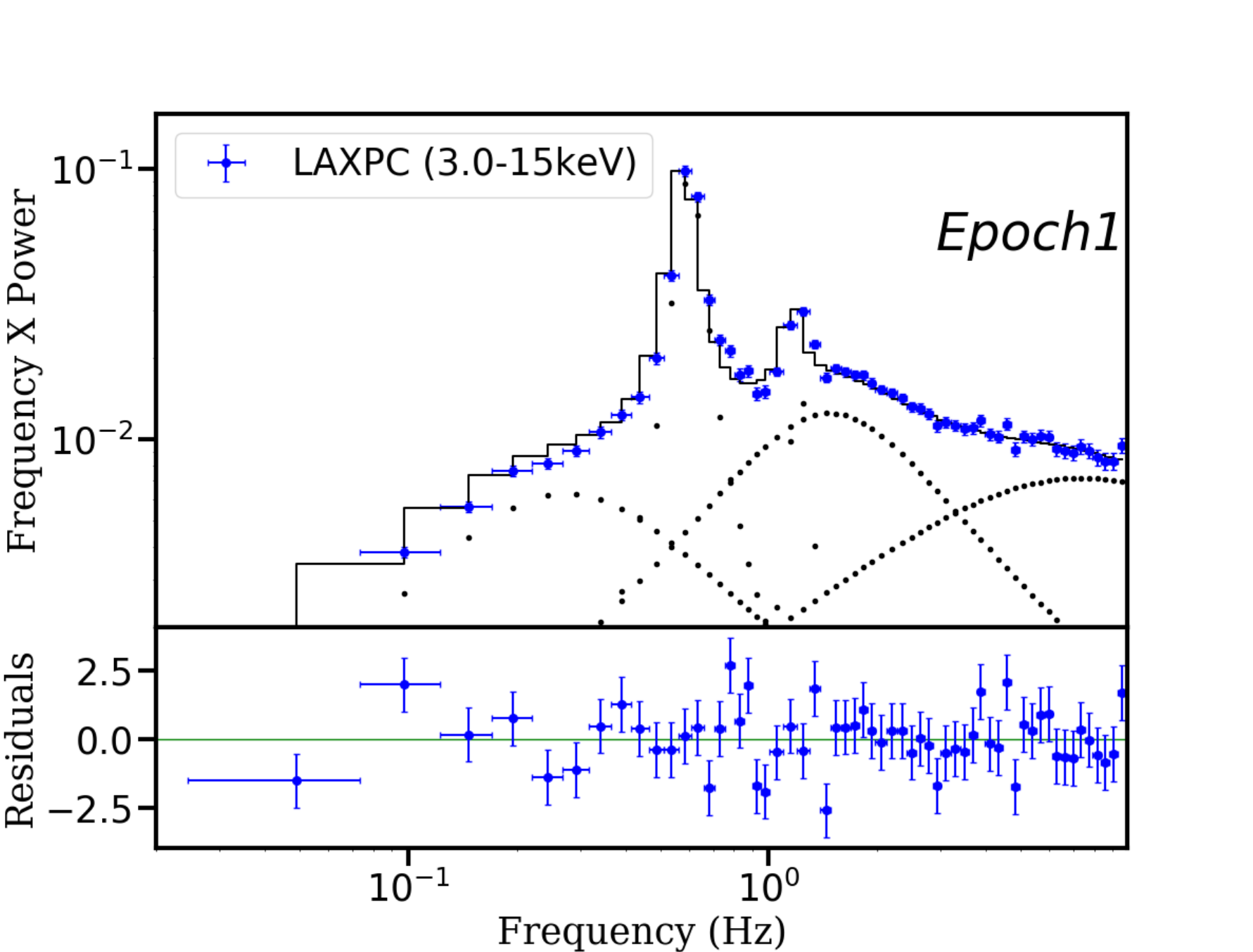}
     \includegraphics[trim=0cm 0cm 2cm 0cm, clip=true, height=6.5cm, width=0.49\linewidth]{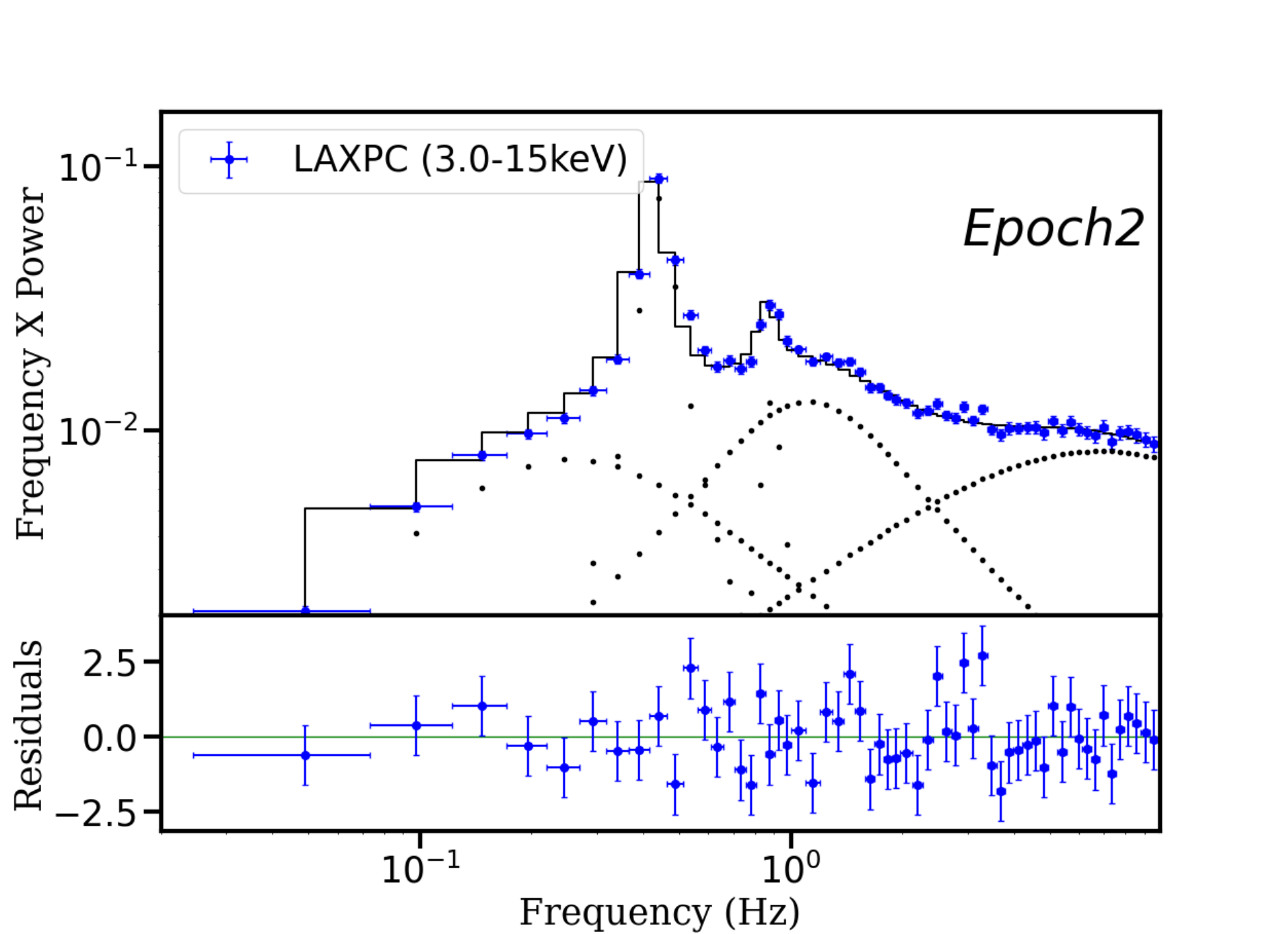}\\
    \caption{The power density spectra of LAXPC is shown in the energy range of 3.0-15 keV for Epoch1 (Left Panel) and Epoch2 (Right Panel). The PDS has been adjusted for Poisson's noise and background, and the fitting was performed with five Lorentzian components. The  lower panel displays the residuals. The PDS reveals distinct and pronounced QPO and their corresponding upper harmonic superimposed on broadband noise.}
    \label{fig:pds_fit}
\end{figure*}

\section{AstroSat Observations and Data Reduction}
\label{sec:Data}

\textit{AstroSat} captured brief observations of \bh\ during its full outburst in 2016 (as reported by \cite{swadeshchand_astro}) and a potentially failed outburst in 2017 \citep{jin2017swift} with its \textit{LAXPC} and \textit{SXT} instruments. The observations were conducted on 2016 March $9^{th}$ from 09:46:02 to 19:06:11 (hereafter, Epoch1) with 11.7~ks of exposure time and on 2017 August $8^{th}$ from 7:42:12 to 18:32:19 (hereafter, Epoch2) with 14.5~ks of exposure time. The \textit{MAXI} lightcurve (2-20~keV) for \bh is shown in Figure~\ref{fig:maxi}. The vertical lines mark the \textit{AstroSat} detection of Epoch1 (blue) and Epoch2 (magenta). We utilize combined \textit{LAXPC} and \textit{SXT} data to study its energy spectrum.

\textit{AstroSat's} \textit{LAXPC} provides a large effective area for energies 3.0-80.0~keV ($\sim6000$~$cm^2$) and high timing resolution of 10~$\mu$s \citep{laxpc_agrawal2017}. Event mode Level 1 data for each epoch was extracted from \textit{AstroSat} archive. Data reduction was performed with \textit{LAXPC} software (version as of May, 2018)\footnote{\url{http://astrosat-ssc.iucaa.in/?q=laxpcData}}. The lightcurve and spectra were obtained using \textit{laxpc\_make\_lightcurve} and \textit{laxpc\_make\_spectra} subroutines. \textit{AstroSat's} \textit{SXT} observes in lower energy X-ray band of 0.3-8.0~keV, with high sensitivity with an effective area of approximately 90~$cm^2$ at $\sim$1.5~keV \citep{sxt_singh2017}. Photon Counting mode Level 2 data for \textit{SXT} was extracted from \textit{AstroSat} archive for each epoch. The reduction of data was done with \textit{SXT} software\footnote{\url{https://www.tifr.res.in/~astrosat_sxt/dataanalysis.html}}. The image of the source was extracted using the merged eventfile and a circular region of radius 15~arcmin centered on the source was considered for further data analysis. Also, latest RMF (sxt\_pc\_mat\_g0to12.rmf) and Background (SkyBkg\_comb\_EL3p5\_Cl\_Rd16p0\_v01.pha) files were used. The modified ARF file (sxt\_pc\_excl00\_v04\_16oct21\_20190608.arf) was corrected for the vignetting effect using the \textit{SXTARFModule} tool.  The spectrum for \textit{SXT} was extracted using the HEASoft (version 6.26.1) package XSELECT. The \textit{SXT} spectrum was further grouped with the background, RMF and corrected ARF files by the interactive command grppha producing the final \textit{SXT} spectrum.

\section{Analysis and Results}
\label{sec:Analysis}

\subsection{Spectral Modelling}
\label{sec:Spectral}

We start with jointly fitting the spectrum extracted from \textit{LAXPC} and \textit{SXT} detectors in the broad energy range of 0.7-50.0~keV. The time-averaged spectra could be well described by two main components; the multicolor blackbody disk emission  and the thermal comptonized emission at higher energies. To model the thermal disk emission we use XSPEC model \textit{Diskbb} \citep{diskbb} which describes the disk in terms of inner disk temperature (\tin) and normalization ($N_{Disk}$). For the high energy powerlaw tail, we use the XSPEC model \nth \citep{nthcomp2} which has \tin, spectral photon index ($\Gamma$), electron temperature ($kT_{e}$) and normalization ($N_{\Gamma}$) as parameters. To be consistent, \tin\ of \nth was tied to that of \textit{Diskbb}. The electron temperature was fixed to a fiducial value of 30~keV, which is in accordance to the value estimated by \cite{kte_30keV_wang20222018} for H~1743-322 during its failed outburst in 2018.
All other parameters were kept free to vary in the fit. For line of sight absorption, we use XSPEC model \textit{TBabs} \citep{tbabs}.
Different values of Hydrogen column density ($n_H$) have been reported for this source therefore we fix it to  2.2 $\times$ $10^{22}$~$ cm^{-2}$ consistent with \cite{mcclintock2009,miller2006nh,corbel2006nh,parmar2003nh,shidatsu2014nh}. 
A 3$\%$ systematic error is also added to the spectral fitting. We fit the spectra with  model \textit{TBabs*(Diskbb+Nthcomp)} and found it to be a good fit for both epochs, the best fit values of parameters are reported in upper panel of Table \ref{tab:Spectral_fits}.  The spectral fitting suggests that both epochs were in a roughly similar spectral state with high inner disk temperature of $\sim$1~keV and powerlaw index of $\sim$1.6 consistent with that obtained earlier by \cite{swadeshchand_astro,mcclintock2009,chen2010}.

\renewcommand\arraystretch{1.5}
\begin{table}
\centering
\caption{Best fit values of spectral parameters found by joint fit of LAXPC and SXT in the energy range of 0.7-50~keV with model TBabs$\times$(Nthcomp+Diskbb) and TBabs$\times$(Thcomp$\otimes$Diskbb) for both epochs. The reported errors are given at a 90$\%$ confidence level.}

\begin{tabular}{c|ccc} \hline 
\textbf{Model} & \textbf{Parameter} & \textbf{Epoch1} &  \textbf{Epoch2}\\
\hline
\textit{{Const}} &  & $0.93^{+0.04}_{-0.04}$ & $1.07^{+0.04}_{-0.04}$ \\
\textit{{TBabs}} & $n_H$~($10^{22}$ $cm^{-2}$) & $2.2(f)$ & $2.2(f)$ \\
\textit{{Nthcomp}} & $\Gamma$ & $1.64^{+0.02}_{-0.02}$ & $1.61^{+0.02}_{-0.02}$ \\
& $N_{\Gamma}$ & $0.13^{+0.02}_{-0.01}$ & $0.11^{+0.01}_{-0.01}$ \\
\textit{{Diskbb}} & $kT_{in} (keV)$ & $1.00^{+0.08}_{-0.098}$ & $1.04^{+0.08}_{-0.11}$ \\
& $N_{Disk}$ & $18.46^{+2.56}_{-3.15}$ & $12.39^{+1.99}_{-1.64}$ \\
\hline 
\textit{{$\chi^2/dof$}}  &  & 105.36/116 &  85.62/115\\
\hline 
\textit{{Const}} &  & $0.94^{+0.04}_{-0.04}$ & $1.08^{+0.04}_{-0.04}$ \\
\textit{{TBabs}} & $n_H$~($10^{22}$ $cm^{-2}$) & $2.2(f)$ & $2.2(f)$ \\
\textit{{Thcomp}} & $\tau$ & $3.51^{+0.06}_{-0.06}$ & $3.63^{+0.06}_{-0.07}$ \\
& $\dot{H}  (keV cm^{-3} s^{-1})$ & $3.03^{+0.05}_{-0.04}$ & $3.03^{+0.08}_{-0.05}$ \\
& $f_{sc}$ & $0.74^{+0.07}_{-0.06}$ & $0.78^{+0.09}_{-0.06}$ \\
\textit{{Diskbb}} & $kT_{in} (keV)$ & $0.97^{+0.07}_{-0.09}$ & $1.01^{+0.09}_{-0.12}$ \\
& $N_{Disk}$ & $64.12^{+22.04}_{-13.08}$ & $49.09^{+23.08}_{-11.19}$ \\
\hline 
\textit{{$\chi^2/dof$}}  &  & 104.74/116 &  86.05/115\\
\hline 
\end{tabular}
\begin{flushleft}
Note: f in parentheses (f) denotes a fixed parameter.    
\end{flushleft}
\label{tab:Spectral_fits}
\end{table}

\renewcommand\arraystretch{1.5}
\begin{table}
\centering
\caption{Best fit values of parameters obtained by fitting the power density spectra in 3-15~keV  with five Lorentzian components. The reported errors are given at a 90$\%$ confidence level.}

\begin{tabular}{c|cccc|} \hline \hline

\textbf{Component} & \textbf{Parameter} & \textbf{Epoch1} & \textbf{Epoch2}  \\

\hline \hline

\textsc{{L1}} & $f_{QPO}$ (Hz) & $0.599^{+0.002}_{-0.002}$ & $0.439^{+0.002}_{-0.002}$  \\
     (QPO)    & $\sigma_1$ & $0.08^{+0.01}_{-0.01}$ & $0.069^{+0.008}_{-0.007}$ \\
               & $N_{L1}~(10^{-2})$ & $2.29^{+0.11}_{-0.11}$ & $2.12^{+0.14}_{-0.13}$ \\
               & $fRMS_{L1}~($\%$)$ & $15.13$ & $14.56$ \\
               \hline
 \textsc{{L2}} & $f_2$ (Hz) & $1.21^{+0.01}_{-0.01}$ & $0.88^{+0.01}_{-0.01}$\\
   (Upper   & $\sigma_2$ & $0.13^{+0.03}_{-0.03}$ & $0.10^{+0.030}_{-0.025}$\\
        Harmonic)       & $N_{L2}~(10^{-2})$ & $0.30^{+0.06}_{-0.06}$ & $0.24^{+0.07}_{-0.06}$ \\
            & $fRMS_{L2}~($\%$)$ & $5.48$ & $4.90$ \\
            
               \hline
 \textsc{{L3}} & $f_3$ (Hz) & $0.15^{+0.01}_{-0.02}$ & $0.11^{+0.02}_{-0.11}$\\
               & $\sigma_3$ & $0.46^{+0.11}_{-0.07}$ & $0.44^{+0.58}_{-0.10}$\\
               & $N_{L3}~(10^{-2})$ & $1.47^{+0.25}_{-0.17}$ & $1.93^{+2.04}_{-0.33}$\\
               & $fRMS_{L3}~($\%$)$ & $12.12$ & $13.89$ \\

               \hline
 \textsc{{L4}} & $f_4$ (Hz) & $1.09^{+0.14}_{-0.13}$ & $0.85^{+0.07}_{-0.16}$\\
               & $\sigma_4$ & $1.98^{+0.13}_{-0.31}$ & $1.40^{+0.38}_{-0.20}$\\
               & $N_{L4}~(10^{-2})$ & $2.31^{+0.16}_{-0.45}$ & $2.30^{+0.58}_{-0.22}$\\
               & $fRMS_{L4}~($\%$)$ & $15.20$ & $15.17$ \\

               \hline
\textsc{{L5}} & $f_5$ (Hz) & $0.0(f)$ & $0.0(f)$\\
              & $\sigma_3 $ & $14.76^{+3.92}_{-2.71}$ & $13.68^{+2.11}_{-1.57}$\\
              & $N_{L5}~(10^{-2})$ & $2.25^{+0.19}_{-0.22}$ & $2.62^{+0.23}_{-0.25}$\\
              & $fRMS_{L5}~($\%$)$ & $15.00$ & $16.19$ \\
              
       \hline     
                           
\textbf{$\chi^2/dof$}  & & \textbf{69.22/45} & \textbf{64.37/45}\\

\hline 

\end{tabular}
\\
\begin{flushleft}
\hspace{2cm} 

\end{flushleft}
\label{tab:pds_fits}
\end{table}

\subsection{Timing Analysis}
\label{sec:Timing}

\subsubsection{Fitting the Power Density Spectrum}

To study the variability of \textit{LAXPC} data we computed the PDS using the LAXPC subroutine \textit{$laxpc\_find\_freqlag$}. The subroutine first generated a lightcurve in 3.0-15~keV band with a time resolution of 0.005~s. The lightcurve is divided into large number of segments ($>500$) and for each segment, PDS is calculated. Subsequently, all PDS are then averaged to produce the final PDS which is deadtime Poissons noise and background corrected. It is normalized as per \cite{rms_norm_miyamoto1992} such that power density is derived in units of $(RMS/mean)^2/Hz$ and its integration gives square of fRMS variability. Following \cite{lorentz_belloni2002unified}, we implement multi-Lorentzian model to fit the PDS and found that five Lorentzian components were required to fit the QPO, its harmonic and broad noise components.  Each Lorentzian is described with three parameters; centroid frequency (f), FWHM ($\sigma$) and  normalization ($N_L$) of respective component. Table \ref{tab:pds_fits} shows the list of best fit parameters with errors estimated in 90$\%$ confidence region along with fRMS ($\%$) contribution for each component which is calculated by taking the square root of $N_L$.  The fitted PDS with individual components is shown in Figure~\ref{fig:pds_fit}.

The best fit parameters of first two Lorentzian components (L1, L2) show that the QPO occurs at a frequency of $\sim$0.60~Hz (Quality factor; Q$\sim$7.5)  with its upper harmonic at $\sim$1.21~Hz (Q$\sim$9.3) for Epoch1 and at $\sim$0.44~Hz (Q$\sim$6.4) with its upper harmonic at 0.88~Hz (Q$\sim$8.8) for Epoch2. Whereas,  rest of the components (L3, L4 and L5) describe the broad noise components in the PDS. Based on significant contributions from hard powerlaw component ($\Gamma$ $\sim1.6$) in the spectra and on the QPO frequency compared to other detections done for its previous outbursts (See e.g. \cite{debnath2013}), the system was in the hard state during both observations.

\subsubsection{Generating fRMS and time-lag spectra}

\begin{figure*}
    \centering{
     \includegraphics[trim=1cm 0cm 0cm 0cm, clip=true, height=5.8cm, width=0.52\linewidth]{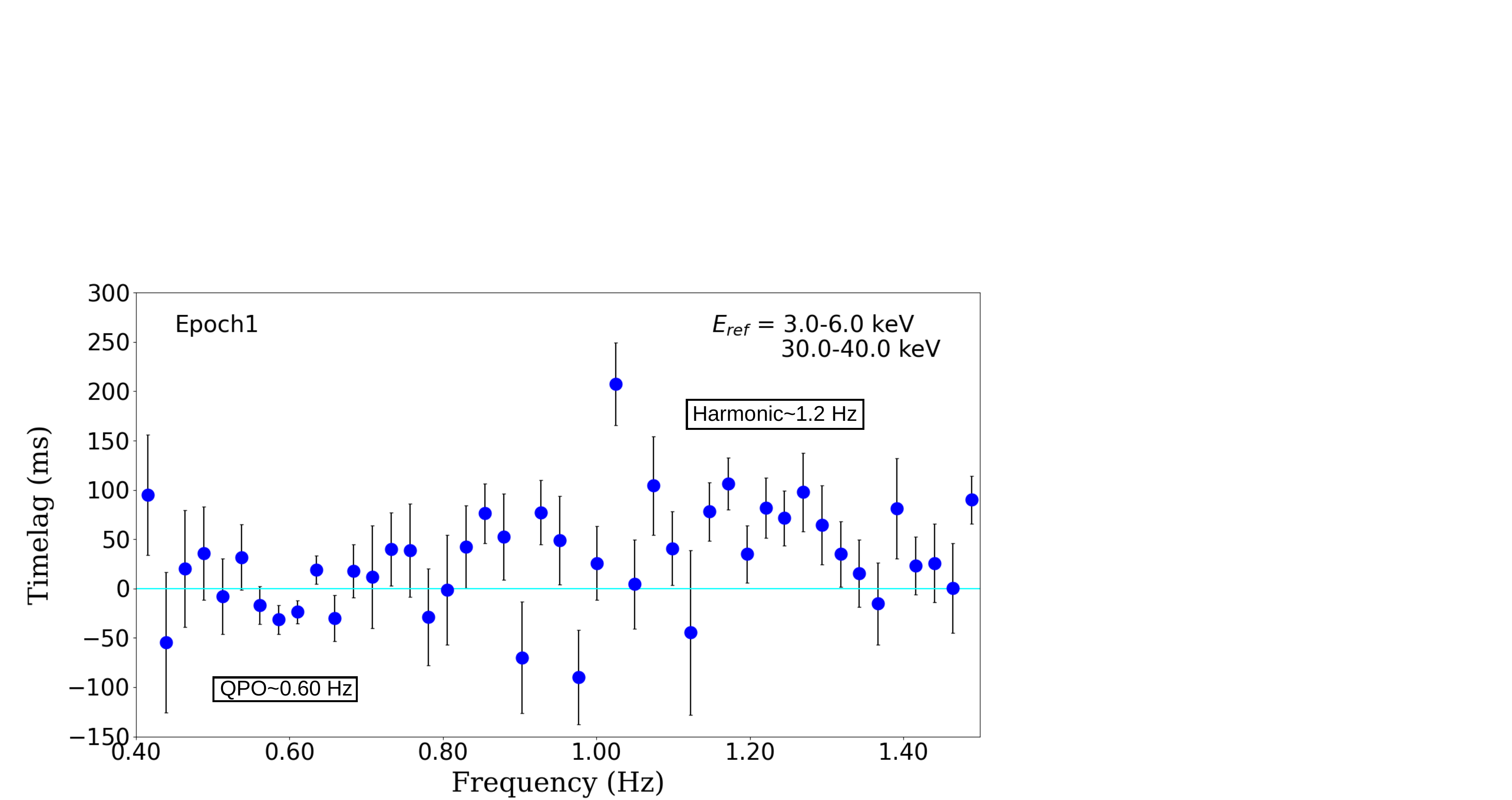}
     \hspace{-1cm}
      \includegraphics[trim=0cm 0cm 2cm 0cm, clip=true, height=5.8cm, width=0.52\linewidth]{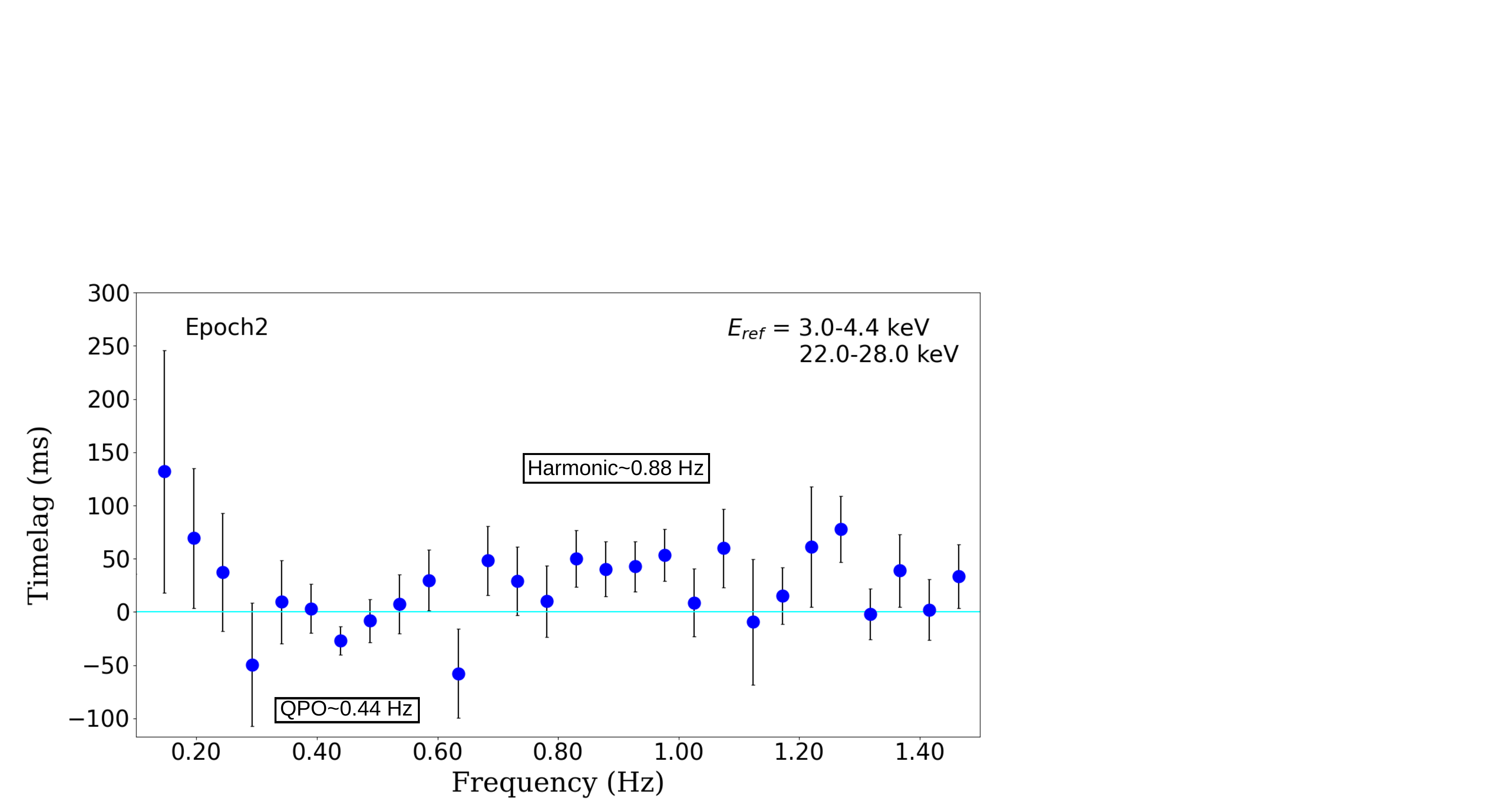}\\
    \hspace{-1cm}
    \caption{Timelag as a function of Fourier frequency for Epoch 1; Left panel and Epoch 2; Right panel. The two energy bands used for calculation of time lags are mentioned on the upper right corner with reference energy denoted as $E_{ref}$. The distinct region for the QPO and upper harmonic frequencies are marked.}
    \label{fig:tl_freq}
    }
\end{figure*}

To explore the energy-dependent nature of the source we next studied the variation of fractional amplitude and time-lags with energy for QPO and the upper harmonic (for details on time-lag calculation, see \cite{lag_nowak1999low}). We again used the subroutine \textit{$laxpc\_find\_freqlag$} to generate the PDS, as described in Section 3.2.1,  in multiple energy bands. To find the fRMS spectra, each PDS was fitted with the best-fit model in Table \ref{tab:pds_fits} with frequency and width fixed for all Lorentzian components, only allowing normalization to vary. The fRMS of the QPO and harmonic were taken as the square root of the normalization of the \textit{Lorentzian} component describing the respective feature. To generate the time-lag spectra, \textit{LAXPC} subroutine provides inputs such as the frequency resolution ($\Delta f$) and frequency at which time-lag has to be computed ($f$). The subroutine first generates PDS in different energy bands, with one of the bands set as reference. It then estimates the phase of the cross correlation function for each energy band with the reference band. The time-lag was calculated by dividing the phase-lag by $2\pi f$. It is important to note that the errors on fRMS has two components, the measurement uncertainties and due to inherent stochastic nature of the lightcurve \citep{vaughan2003}. The stochastic error is due to the finite length of the time segment which may be correlated in different energy bands. Since one of the basic assumptions in $\chi^2$ fitting is that the error bars are uncorrelated, this may lead to reduced $\chi^2$ being significantly less than unity. An elegant solution to this problem is provided in \cite{ingram2019}. By considering the correlation of variability in different energy bands, they have effectively corrected the errors pertaining to power and phase-lag. We have adopted the approach and applied their equations 19 and 20 to correct the errors on phase-lag and power, respectively. To further check the behaviour of time-lag at different frequencies and to see if QPOs and their harmonic do have distinct behavior, we plot the time-lag of hard photons with respect to photons in a reference soft band ($E_{ref}$) as a function of frequency in Figure \ref{fig:tl_freq} for Epoch1 (Left panel) and Epoch2 (Right panel). We observe that the time-lag associated with the QPO (and harmonic) frequency exhibits a clear negative (and positive) time lag during each Epoch.

Obtained fRMS and time-lag spectra corresponding to each epoch is shown in Figure~\ref{fig:fit_rms_lag_2016} and \ref{fig:fit_rms_lag_2017}, where the Left panels show fRMS spectra and Right panels show time-lag spectra. The reference energy band used for calculating the time-lag is taken to be 3.0-6.0~keV for Epoch1 and 3.0-4.4~keV for Epoch2. The fRMS spectra of the QPO exhibits a trend of gradual increase up to 10~keV. We see that upon reaching this threshold in Epoch1, the fRMS initially decreases but then resumes an upward trend, whereas for Epoch2, it consistently declines throughout. As for harmonic, the fRMS decreases with energy and is limited up to 35~keV due to large errorbars above it. As for the time-lag spectra, the soft energy photons lag the high energy photons at the QPO frequency and we see opposite time-lag at harmonic frequencies. Moreover, there is a clear pattern of the time-lag steadily increasing as the energy level increases, with higher energies displaying progressively longer time-lags.

\renewcommand\arraystretch{2.2}\
\begin{table*}
\large
\caption{\large{Best fit values of parameters obtained by fitting the fRMS and time-lag spectra of the QPO with Model2 and of the upper harmonic with Model1.}}
\resizebox{\textwidth}{!}{%
\centerline{\begin{tabular}{p{2.5cm}|p{2.3cm}|p{2.2cm}|p{2.2cm}|p{2.2cm}} \hline 
\centering
& \textbf{Model} & \textbf{Parameter} &  \textbf{Epoch1} & \textbf{Epoch2} \\

\hline 
& & $|\delta kT_{in}|$  & $0.045^{+0.002}_{-0.002}$ & $0.044^{+0.002}_{-0.002}$ \\
& & $|\delta \dot{H}|$  & $0.260^{+0.007}_{-0.007}$ & $0.342^{+0.010}_{-0.010}$ \\
& & $|\delta f_{sc}|$  & $0.041^{+0.013}_{-0.013}$  & $0.043^{+0.009}_{-0.009}$ \\
\textbf{QPO} & \textbf{Model2} & $\phi_{\dot{H}}$ & $-0.023^{+0.027}_{-0.026}$  & $-0.074^{+0.025}_{-0.025}$\\
& & $\phi_{f_{sc}}$ & $+0.36^{+0.29}_{-0.27}$ & $-0.07^{+0.18}_{-0.18}$\\
& & $T_{\dot{H}}$ (millisec) & $6.2\pm7.2$ & $26.8\pm9.1$ \\
& & $T_{f_{sc}}$ (millisec) &  $97.8\pm76.1$ & $25.3\pm65.2$ \\

\hline
 & \textsc{{$\chi^2/dof$}} & & \textbf{1.02 (14.28/14)}  & \textbf{1.48 (16.31/11)}  \\
\hline

&  & $|\delta kT_{in}|$  & $0.0191^{+0.0010}_{-0.0010}$ & $0.0173^{+0.0009}_{-0.0009}$\\
\textbf{Upper} & \textbf{Model1}  & $|\delta \dot{H}|$  & $0.084^{+0.013}_{-0.013}$  & $0.113^{+0.021}_{-0.021}$\\
\textbf{Harmonic} & & $\phi_{\dot{H}}$  & $+0.402^{+0.083}_{-0.077}$  &  $+0.330^{+0.089}_{-0.077}$\\
& & $T_{\dot{H}}$ (millisec) & $52.2\pm10.4$ & $59.7\pm15.0$ \\

\hline
 & \textbf{$\chi^2/dof$} & & \textbf{1.12 (11.16/10)}  &  \textbf{0.93 (9.27/10)}\\
\hline \hline
\multicolumn{5}{l}{%
  \begin{minipage}{14.5cm}%
     \large Note: In general,  $|\delta A|$ represents fractional variation of parameter A. $\phi_A$ represents phase lag of $\delta A$ with respect to $|\delta kT_{in}|$ and $T_A$ is the corresponding time-lag in milliseconds.%
  \end{minipage}%
}
\end{tabular}}}
\label{tab:rms_lag_fit}
\end{table*}

\subsubsection{Modelling the energy-dependent properties}

Here, we briefly discuss the technique developed by \citetalias{akash_garg2020identifying} to model the fRMS and time-lag spectra of the peaked components using spectral information.  They developed the model considering a simple picture of the accretion flow where in hard state the thin and optically thick disk is truncated at a certain radius far from the ISCO and is converted into a hotter, optically thinner medium of high energy electrons known as hot inner flow or corona. Physics behind this conversion of disk into the hot medium is yet not clear but is sometimes associated with evaporation of the disk in inner regions \citep{disk_evaporation_meyer1994accretion}. As discussed in Section 3.1, the spectra is dominated by two components which in this geometry could be explained such that the disk photons generated due to thermal emission can interact with the high energy electrons and get Compton up-scattered leading to a high energy tail. The model by \citetalias{akash_garg2020identifying} considers an intrinsic origin of the QPO corresponding to variations in physical spectral parameters of the accretion flow components. As per their technique,  first order variations in spectral parameters will induce change in the steady state spectrum F(E), as $\Delta F(E)=\sum_{j=1}^N \frac{\partial F(E)}{\partial \alpha_j} \Delta \alpha_j$ here $\alpha_j$ is the spectral parameter with j varying from 1 to N which is the total number of parameters and $\Delta \alpha_j$ is the variation introduced in the $j^{th}$ parameter. The first order derivatives are then numerically calculated by the code which are further used to model fRMS, defined as $|\Delta F(E)|/F(E)$ and phase lag which is the argument of ${\Delta F(E_{ref})}^*\Delta F(E)$, here $E_{ref}$ is the reference energy band used to compute the cross spectra. Further attributes of the model are discussed in detail in \citetalias{akash_garg2020identifying,garg2022}.

We utilize their model to fit timing variability of our system \bh. The spectral fitting with model \textit{Diskbb} and \textit{Nthcomp} is already discussed in Section 3.1.  We now fit the spectra by replacing \nth with \textit{Thcomp}, which is an improved version of \nth model. It is a convolution component and describes the continuum shape of thermalized emission better by assuming spherical distribution of thermal electrons \citep{thcomp_zdziarski2020}. It has similar parameters to \nth with addition of covering fraction ($f_{sc}$) indicating fraction of seed photons getting comptonized. It also allows one to choose Thomson optical depth ($\tau$) as a fit parameter instead of $\Gamma$. The model \textit{Thcomp} has been further modified by \citetalias{garg2022} such that instead of electron temperature it calculates heating rate of corona ($\dot{H}$) iteratively as a model parameter, details of which are given in their work (see \textit{eq. 2} of \citetalias{akash_garg2020identifying}). We implement this modified model to the joint spectrum of \textit{LAXPC} and \textit{SXT}, the corresponding best fit values are listed in lower panel of Table \ref{tab:Spectral_fits} along with $\chi^2/dof$. The model fitting and the residuals are shown in Figure~\ref{fig:spectral_fit}. Both epochs show similar value of parameters within errorbars, with a high fractional scattering of $\sim$0.76  and an optical depth of $\sim$3.57. To check for the presence of Fe k$\alpha$ line emission we add a \textit{Gaussian} in region 6.4-7.0~keV and find that the fit improves by $\Delta\chi^2$ $\sim$20 for Epoch1 and $\sim$3 for Epoch2. However the strength of the line is weak with equivalent widths of $\sim$0.36 and $\sim$0.13~keV for the two Epochs and hence is not expected to contribute significantly to the timing properties. Since the model used here to fit the timing properties assumes that the spectra can be represented only by a disc emission and a thermal Comptonization component, we exclude the line emissions from the analysis. We note that the best fit spectral parameters do not change significantly when the lines are omitted.

\begin{figure*}
    \hspace{-1.2cm}
    \includegraphics[trim=0cm 0cm 0cm 0cm, clip=true, height=5.5cm, width=9cm]{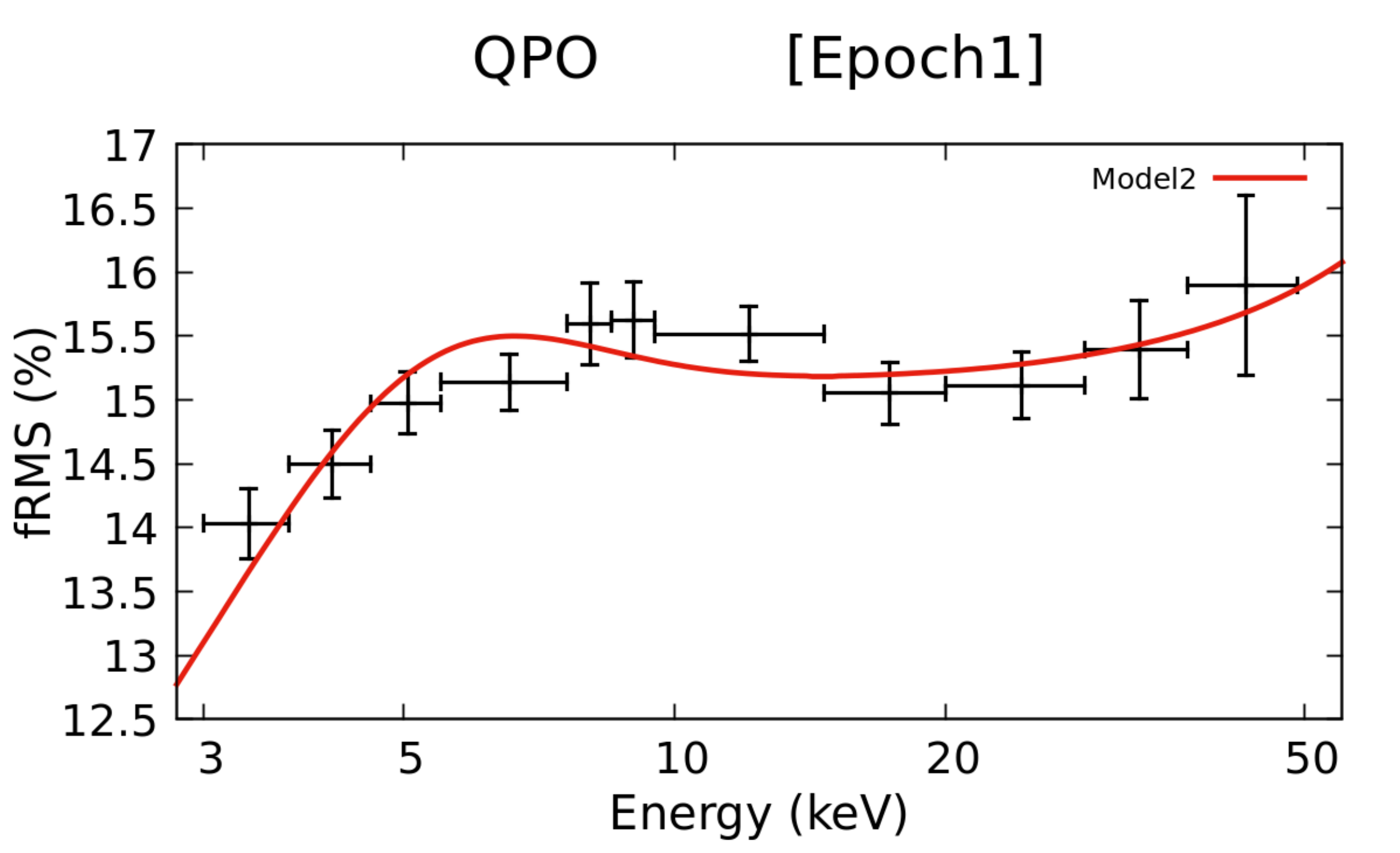}
    \hspace{0cm}
     \includegraphics[trim=0cm 0cm 0cm 0cm, clip=true, height=5.5cm, width=9cm]{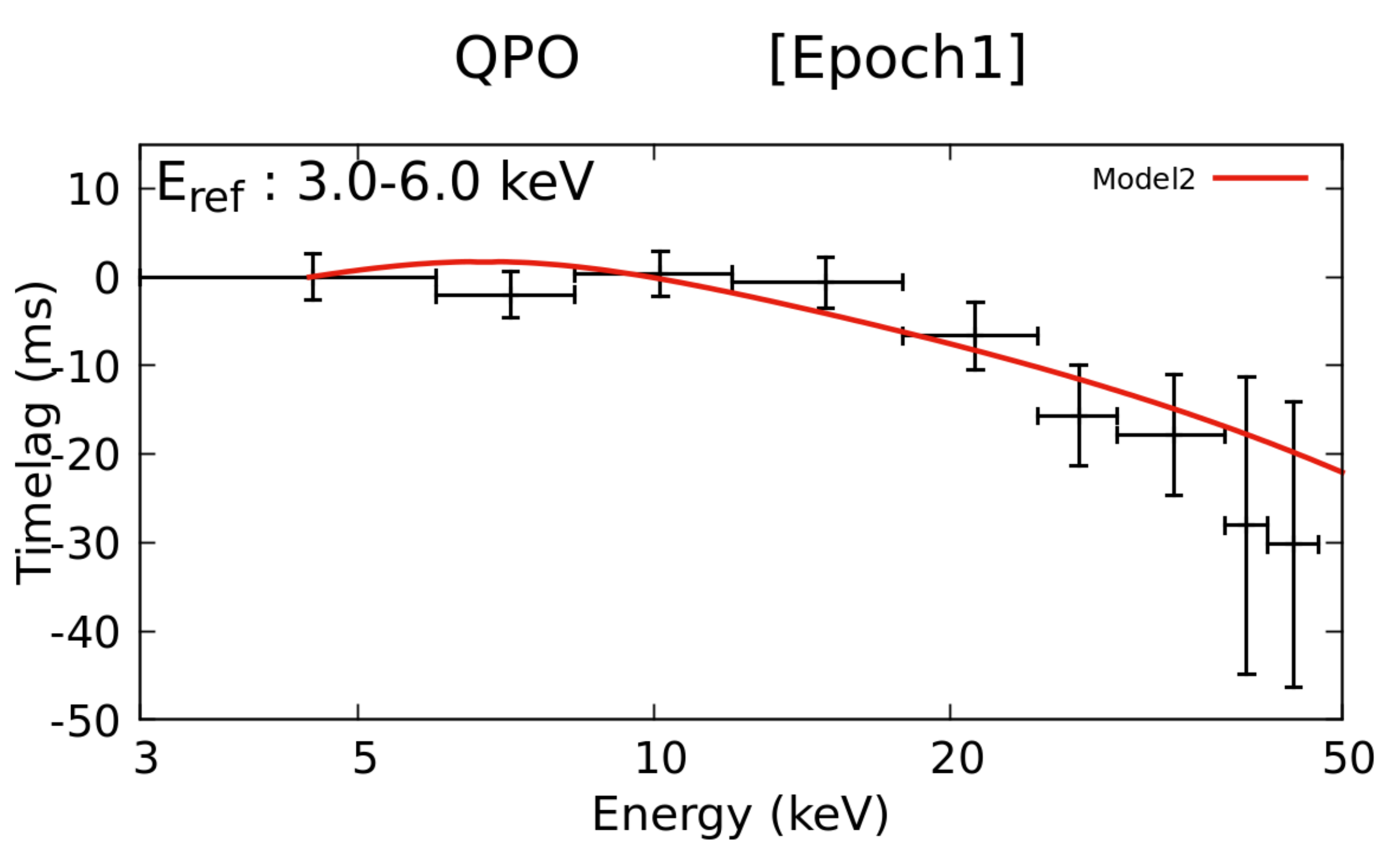}\\
      \hspace{-1.2cm}
      \includegraphics[trim=0cm 0cm 0cm 0cm, clip=true, height=5.5cm, width=9cm]{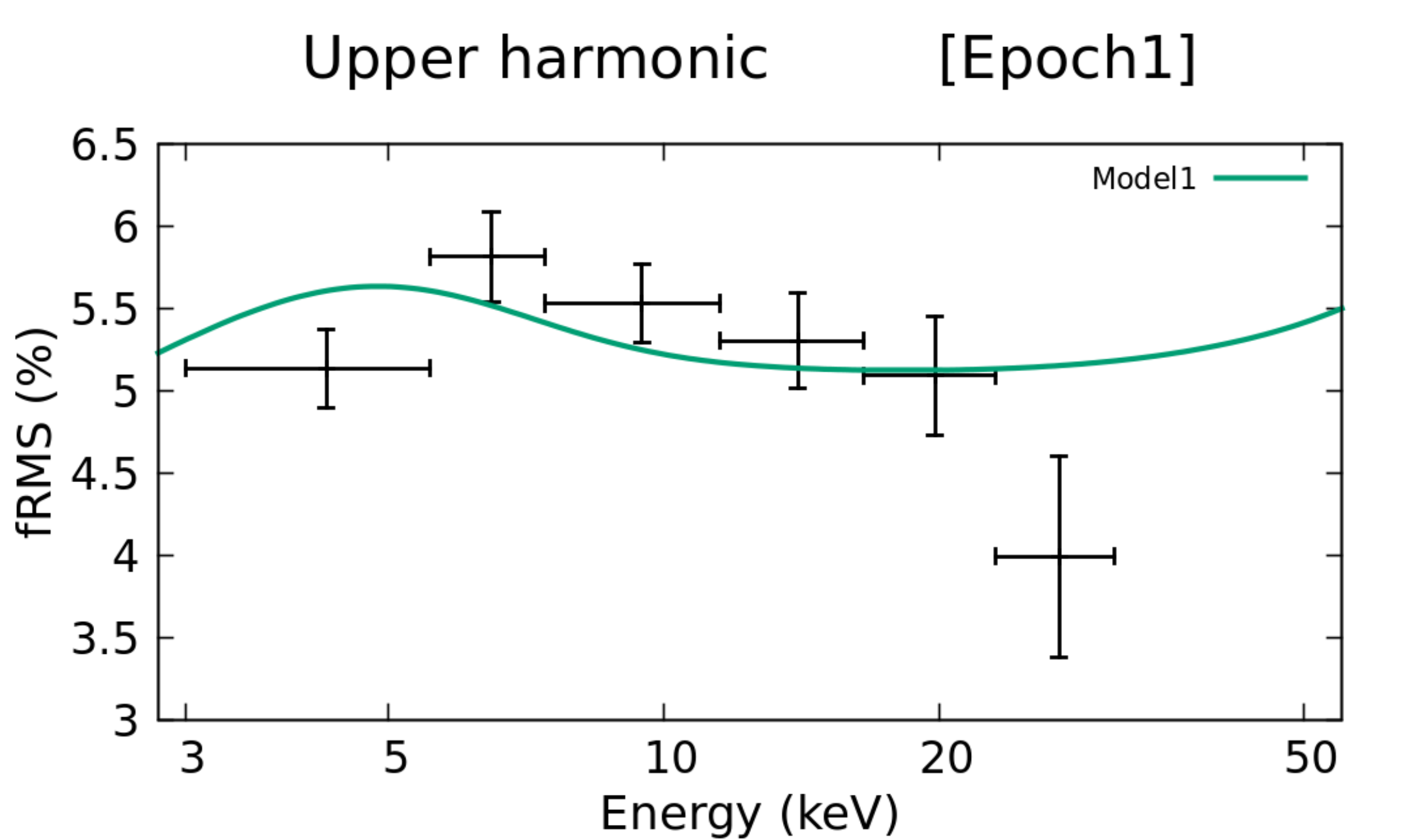}
      \includegraphics[trim=0cm 0cm 0cm 0cm, clip=true, height=5.5cm, width=9cm]{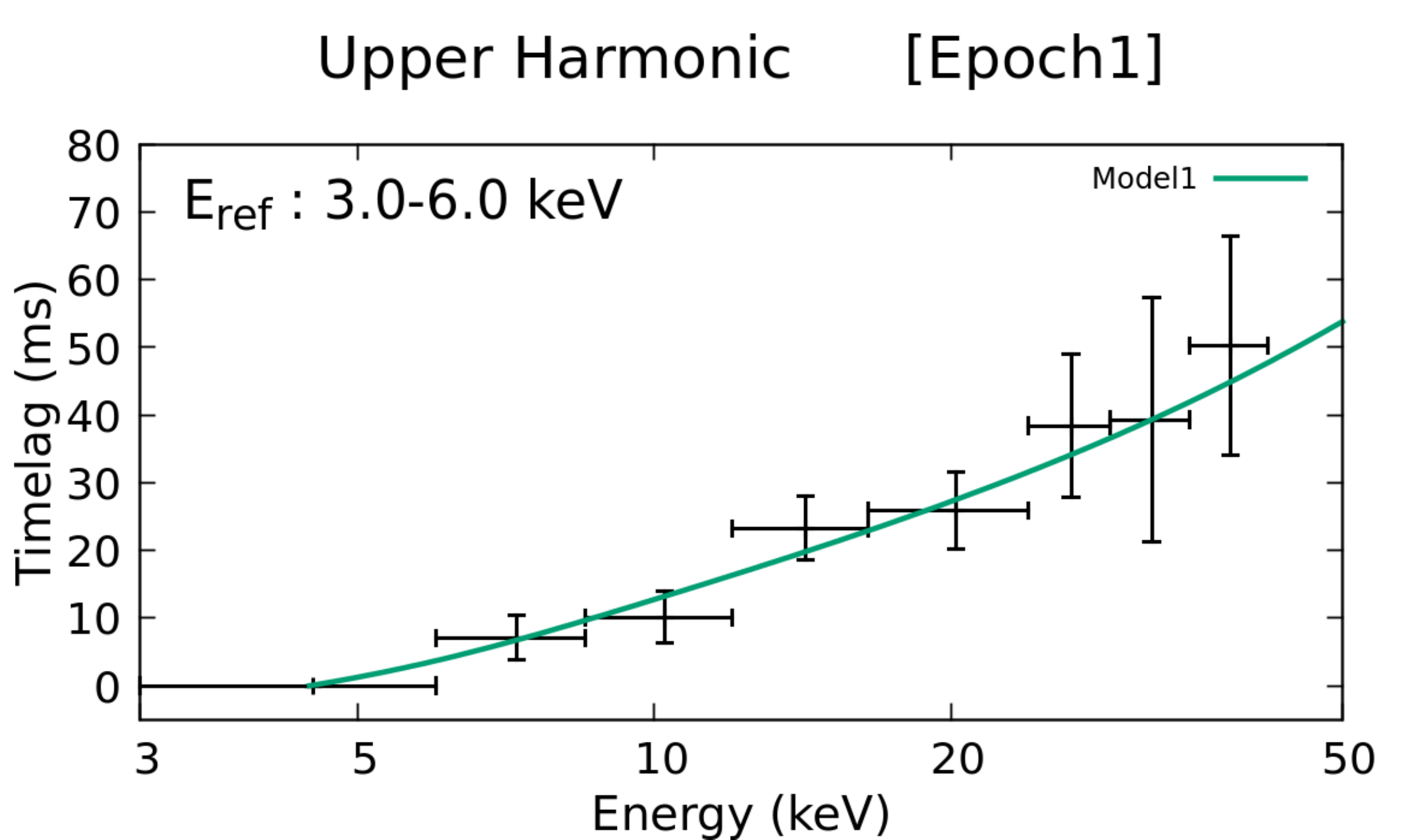}
    \caption{The left panel displays the fractional RMS spectrum, while the right panel shows the time-lag spectrum for Epoch1. The fitting of both spectra for the QPO was conducted using Model2 shown in upper panel, and for the upper harmonic Model1 was utilized as shown in lower panel. The reference energy band used for calculating the time-lag spectra is taken to be 3.0-6.0~keV}
    \label{fig:fit_rms_lag_2016}
\end{figure*}

\begin{figure*}
    \hspace{-1.2cm}
    \includegraphics[trim=0cm 0cm 0cm 0cm, clip=true, height=5.5cm, width=9cm]{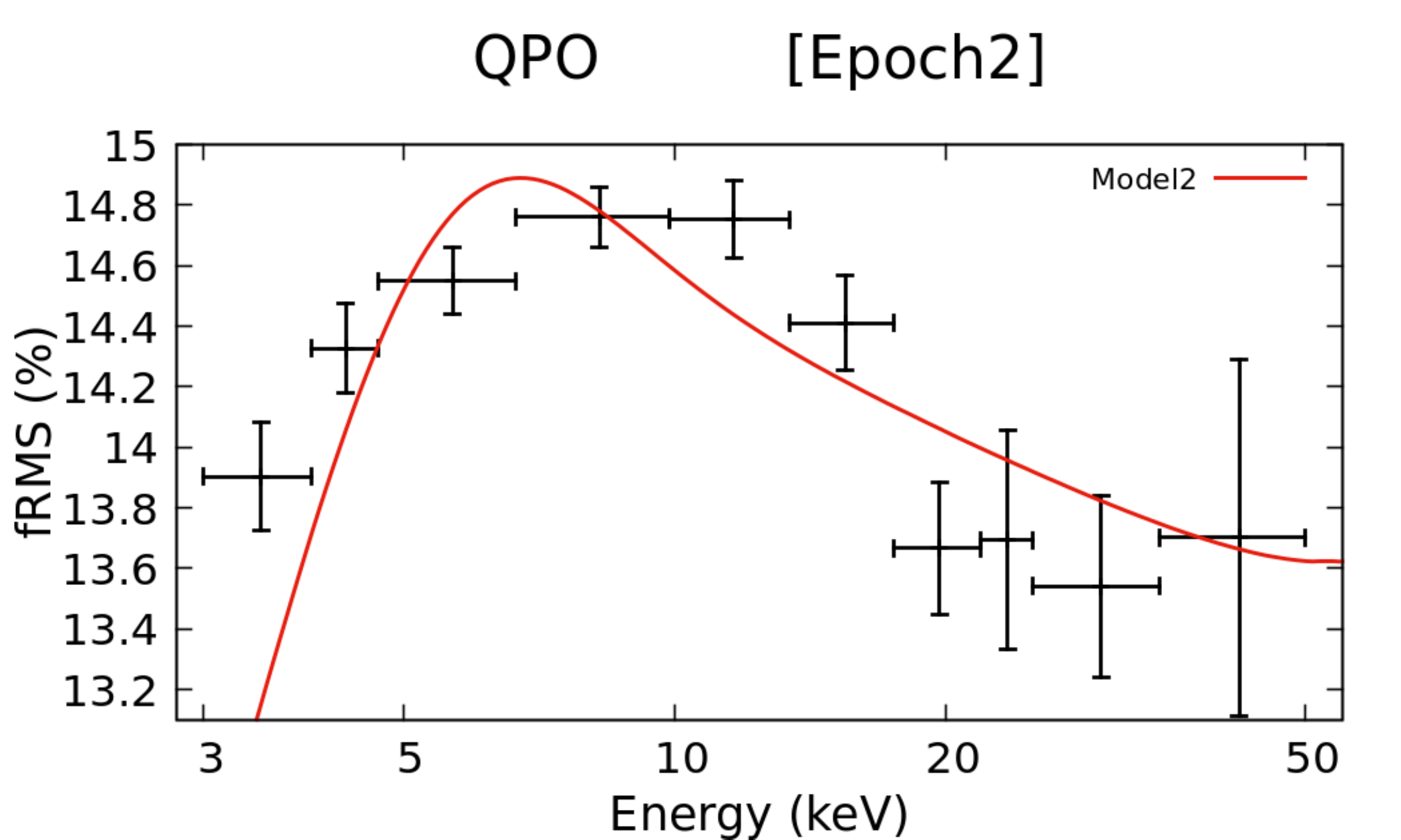}
     \includegraphics[trim=0cm 0cm 0cm 0cm, clip=true, height=5.5cm, width=9cm]{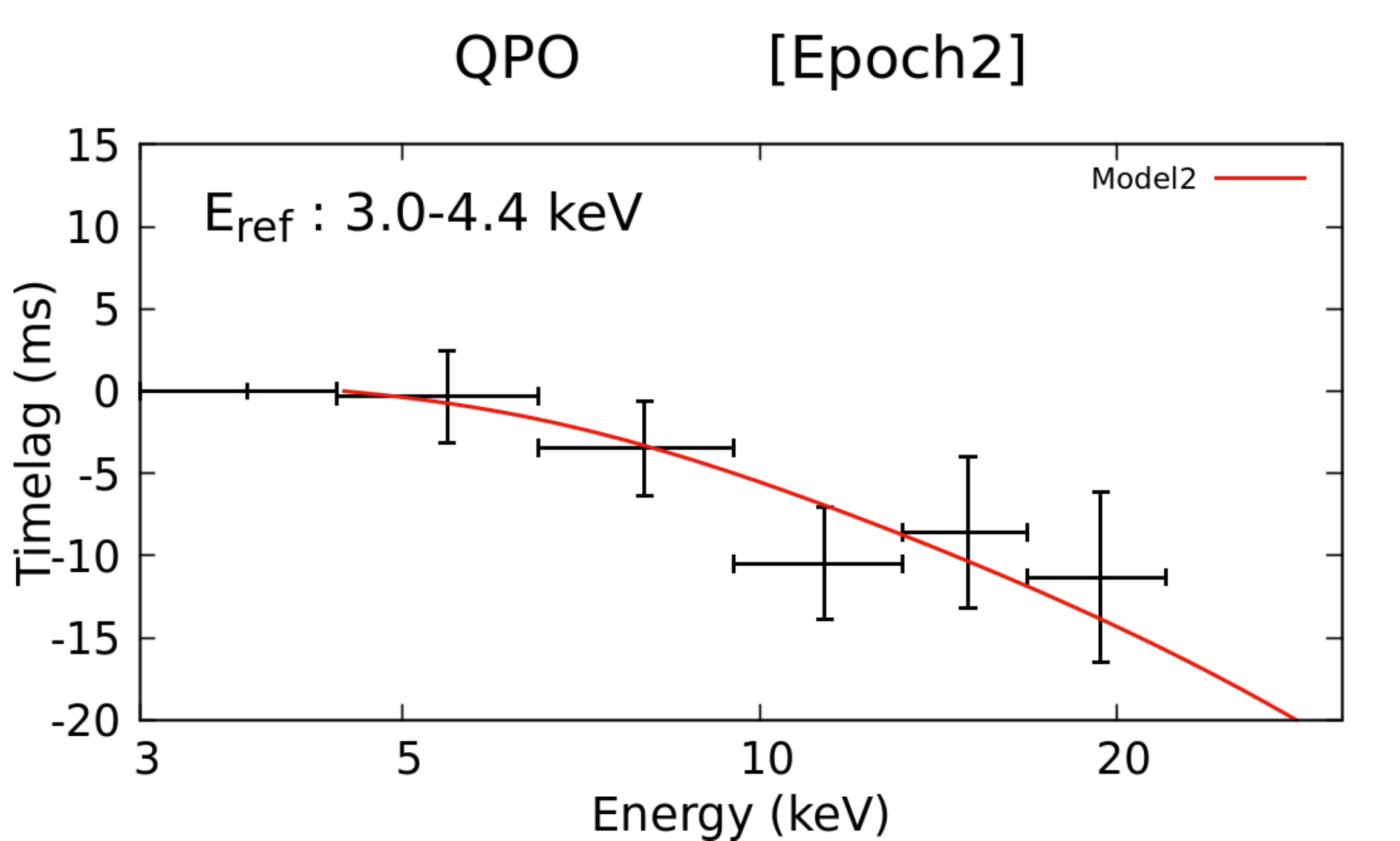}\\
      \hspace{-1.2cm}
     \includegraphics[trim=0cm 0cm 0cm 0cm, clip=true, height=5.5cm, width=9cm]{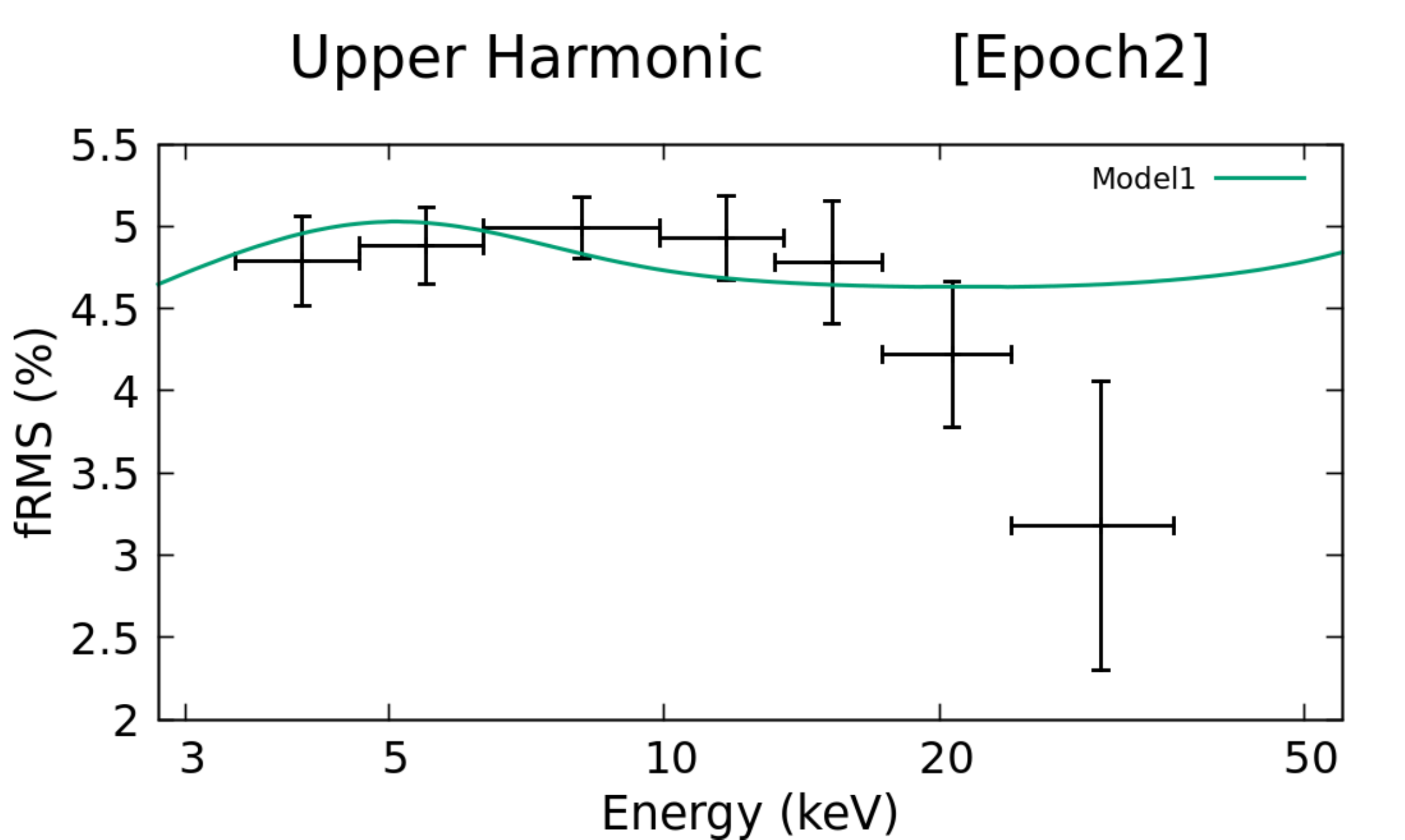}
     \includegraphics[trim=0cm 0cm 0cm 0cm, clip=true, height=5.5cm, width=9cm]{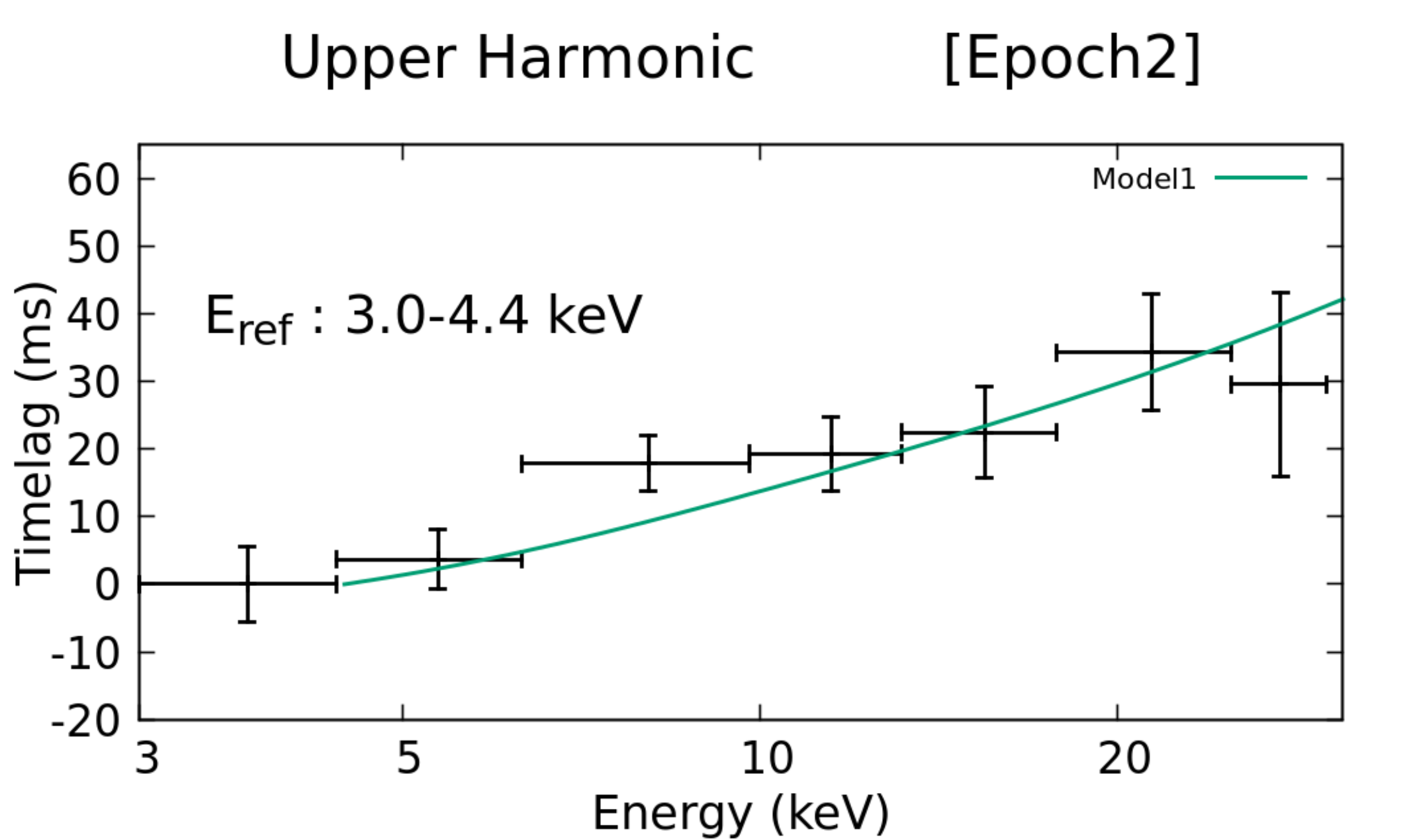}
     \caption{The left panel displays the fractional RMS spectrum, while the right panel shows the time-lag spectrum for Epoch2. The fitting of both spectra for the QPO was conducted using Model2, shown in upper panel, while for the upper harmonic Model1 was sufficient, shown in lower panel. The reference energy band used for calculating the time-lag spectra is taken to be 3.0-4.4~keV.}
    \label{fig:fit_rms_lag_2017}
\end{figure*}

We start by fitting the temporal properties of QPO and its upper harmonic of Epoch1 individually. We first allow small amplitude variations in inner disk temperature ($|\delta kT_{in}|$) and heating rate (|$\delta \dot{H}$|) with \hr varying with a phase lag ($\phi_{\dot{H}}$) with respect to $kT_{in}$, hereafter denoted as Model1 with total of three free parameters. The spectral parameters are fixed at the values found with \textit{Thcomp} model fitting for each epoch. We fit the observed fRMS and time-lag variation with the model and found it to be an ill fit with $\chi^2/dof$$\sim$2.7. Therefore, we additionally introduced variation in fractional scattering ($|\delta f_{sc}|$)  having a phase lag ($|\phi_{ f_{sc}}|$) with respect to $|\delta kT_{in}|$, which we now refer to as Model2 with total of five free parameters and find it to be much better description of the data with $\chi^2/dof$$\sim$1.0. As for the upper harmonic, Model1 fits the variability well with $\chi^2/dof$$\sim$1.1. The list of best fit parameters is given in Table \ref{tab:rms_lag_fit}. The model fitting is shown in Figure~\ref{fig:fit_rms_lag_2016}, upper panel is for QPO and lower panel is for harmonic. For Epoch2, Model1 was also incapable to fit the temporal properties of the QPO with high $\chi^2/dof$$\sim$6.5, therefore following fitting of Epoch1, we fit it with Model2 which provided a much better description with $\chi^2/dof$$\sim$1.48. As for the harmonic, Model1 can describe the properties well with $\chi^2/dof$$\sim$0.93. The list of best fit parameters is given in Table \ref{tab:rms_lag_fit}. The model fitting is shown in Figure~\ref{fig:fit_rms_lag_2017}, upper panel is for QPO and lower panel is for harmonic. The analysis reveals that the variation responsible for producing  the QPO is initially introduced in the heating rate, as indicated by the negative value of $\phi_{\dot{H}}$, which is followed by changes in the inner disk temperature.  This stands in contrast to the behavior observed for the harmonic, where the variability first originates in the inner disk temperature and subsequently affects the heating rate, as indicated by the positive value of $\phi_{\dot{H}}$. For both epochs, while the reduced $\chi^2$ is <2, visual inspection of the fitting reveals that the peak of the fRMS spectra is slightly shifted. In order to get a better description, we attempted, without much success, to vary slightly the time averaged spectral parameters from their best fit values. However, we refrained from exploring models with a larger number of variable parameters and instead have used Model2 as an adequate description of the temporal properties.

\section{Discussion and conclusion}

We conducted a spectral and timing study of two AstroSat observations of \bh extracted near the peak of its outburst in 2016 and 2017. The PDS of each observation displayed sharp QPOs, accompanied with a harmonic component with twice the fundamental frequency. We generated the fRMS and time-lag spectra for each component and found that QPO exhibited soft lags while its corresponding upper harmonic had hard lags. For better understanding of the radiative mechanism behind the variability, we modelled the energy-dependent fRMS and time-lag  of the QPO and its upper harmonic with the model put forward by \citetalias{akash_garg2020identifying}. According to the model, the energy-dependent properties of the variability can be well described with correlated variations of certain spectral components with respective time-delays. Using their general scheme by incorporating variations in $kT_{in}$, $\dot{H}$, $f_{sc}$ with phase lags (Model2), we were able to fit the fRMS and time-lag spectra of QPO. In the case for the harmonic, only variations in $kT_{in}$ and $\dot{H}$ were required (Model1).

In \citetalias{akash_garg2020identifying}, the authors successfully describe the energy-dependent properties of Type-C QPOs of GRS 1915+105, for its AstroSat data of 2017, by introducing variation in heating rate, inner disk temperature, optical depth and disk normalization with corresponding phase lags with respect to heating rate. In their description, they consider a physical scenario in which the variation is initially induced in the heating rate, leading to variation in optical depth. This variation further propagates outwards and drives variation in $N_{Disk}$, which corresponds to a change in the inner radius, and ultimately resulting in variation of $kT_{in}$. In their recent work, \citetalias{garg2022} analysed a long term observation of another BHB system MAXI 1535-571 in its outburst of 2017. To describe the energy-dependent properties of the QPO (Type-C 1.7 to 2.8~Hz), they have changed their model parameters such that instead of $|\delta kT_{in}|$ they use variation in accretion rate $|\delta \dot{M}|$ and in place of $|\delta N_{Disk}|$ they consider ratio of inner radius and accretion rate variation $|\delta R_{in}|/ |\delta \dot{M}|$. 
They successfully use the model to describe the fRMS and time-lag spectra of the QPO by allowing small variations in $\dot{M}$ and $R_{in}$ simultaneously and in $\dot{H}$ with a phase lag with respect to the two parameters. In line with their interpretation, our fitting with Model2 suggests that the variability responsible for the QPO is first generated in the hot inner flow, causing changes in its coronal structure, before propagating outwards and influencing the thermal disk. This is indicated by the variation in the heating rate preceding the variation in either the inner disk temperature or the fractional scattering. In contrast, for the harmonic, the variability is first generated in the thermal disk before propagating inwards towards the hot inner flow. Therefore, within the context of this study, we describe hard lags as when variations in the disc propagate inwards to the corona and soft lags as when the coronal fluctuations propagate outwards to the disc. It should be mentioned that although, stochastic variations are known to propagate inwards due to viscous diffusion \citep{lyubarskii1997},  these variations may also propagate outwards leading to soft lags where the outer region reacts to fluctuations from the inner region (as shown by \cite{mushtukov2018propagating,mummery2023} using Green function approach). Moreover, if these perturbations are carried as sound waves \citep{misra1999}, then they may travel either inwards or outwards depending on their origin.

To be more specific, in this interpretation, the QPO arises due to correlated variations in the accretion rate at the truncated radius and the heating rate of the corona, with a time difference between them. This is partly motivated by spectral fitting results that the QPO frequency corresponds to the dynamical time-scale at the truncated radius \citep{misra2020} and has a dependence on both the value of the radius and accretion rate. This interpretation differs from the model of \cite{origin_ingram2009low} in which the QPO is due to relativistic precession of the inner corona, which is supported by numerical simulations showing the whole inner corona precesses coherently roughly at the lense-thirring frequency of the truncated disc. The precession model also predicts that the reflection features (especially the centroid energy of the Fe fluorescence line) should depend on the precession angle of the inner corona and hence should be a function of the QPO phase. Study of this dependence has been attempted with indications of its presence \citep{ingram2016,ingram2017,nathan2022} at a confidence level of 3.7 sigma. On the other hand, the dependence found of the QPO frequency on both the truncated radius and the accretion rate (\cite{misra2020}; \cite{rawat2023testing}) indicates rather a more complex behaviour. So if the inner corona was not precessing i.e. the QPO is due to heating rate changes, then corresponding coronal structural changes could lead to obscuration of the reflecting disc and hence phase-dependent variation of the Fe line emission. However, such a scenario will perhaps not produce systematic variation in the Fe line centroid energy.

Another interesting method to possibly discern between models, is to examine if the energy-dependent fRMS and time-lag of a QPO depends on the inclination angle of the source. The idea here is that a precessing corona would give rise to different QPO features for different inclination angles while such a dependence will not be seen (or less so) for non-precessing models. Such an analysis is challenging, since energy-dependent fRMS and time-lag can be reliably obtained for few sources and the inclination angle is uncertain for most sources. Taking the above into account, such an analysis has been undertaken which indicates that there is inclination dependence of the QPO properties \citep{van2016inclination,schnittman2006precessing,motta2015geometrical}. A particularly striking result is seen in \cite{van2016inclination} in which for QPO with centroid frequencies $\sim$6~Hz (see their Figure~4), the energy-dependent time-lag reverses sign (with the same magnitude) for high and low inclination angle sources. While, this is based primarily on three sources, the statistical significance of the result is high (99$\%$ for QPOs with frequencies greater than 1~Hz). Although favouring the precession disc model, it is not clear why there should be sign reversal instead of a continuous variation with inclination angle. It should be noted that even for non-precessing model as considered here, the time-lag and fRMS can have inclination angle dependence. Contribution of the disc emission to the corona will have inclination dependence and Comptonized spectra will depend on the viewing angle. This systematic spectral variation will change the contribution of the spectral components for a particular energy band, leading to differences in the fRMS and time-lag. However, this is unlikely to result in a sign reversal for the time-lags. Larger sample study is required to constrain the actual functional dependence of the time-lag with inclination angle in order to ascertain the nature of the QPO.

The energy-dependent analysis undertaken here suggest that the accretion rate at the inner radius oscillates at the QPO frequency. This may seem to imply that both the disc and corona are involved in the phenomenon and that the QPO is not solely due to variations of the inner flow.  However, we note that this is a model dependent statement and cannot be generalized to argue against  those interpretations where the inner flow alone is associated with the QPO phenomenon. For the data presented in this work,  it maybe possible to invoke a model where the energy dependent RMS and time-lag are explained by precession of an hot inner flow. Such a model may need to take into account the reflection  and reprocessing of the hard X-ray emission on the disc and may need to fit the energy spectra with a different model than the one used in this work. Such an analysis could  be similar to the phase resolved spectroscopy  undertaken by \cite{ingram2016} or a simpler variant which predicts the energy dependent RMS and time-lag. An important point here is that in case when reflection or reprocessing in the disc is necessary for the modeling of the RMS at certain energies, then one would expect soft lags at those energies. Optimally such an analysis should be done with broad band data which should also cover lower energies $<$ 4 keV.  Indeed, some NICER studies, such as \cite{rmslowenergy2023} and \cite{alabarta2022variability}, report that for energies <4 keV, the fRMS decreases with energy and that the time-lags are soft, which seems to be consistent with the precession model,  however, detailed modelling is required with wider energy coverage. Therefore wide-band spectral and rapid timing analysis, possible using simultaneous NICER and AstroSat data, may be an important test to differentiate between interpretations, and may enhance our understanding of the phenomena. We also checked if the time delays recorded in our work could be due to the light-crossing time,= R/c, here R is the size of the system and c is the speed of light. Corresponding to the time-lag of heating rate and fractional scattering with respect to inner disk temperature which is roughly 6~ms ($T_{\dot{H}}$) and 98~ms ($T_{f_{sc}}$) for Epoch1, we estimate the size of the system to be respectively at 126 and 1980~$R_g$ for a black hole of mass 10~$M_{\odot}$ which is rather large for the size of the corona, thereby making it unrealistic to be related to the light-crossing time.

The advantage of this approach, as demonstrated here, is that it presents a simplistic picture of distinguishing radiative component that causes variability in the flux, which are then observed as QPOs. By using this method, we were able to distinguish certain physical parameters of the flow (inner disk temperature, fractional scattering and heating rate) whose oscillation could generate the QPOs and we achieve this by fitting the temporal behaviour at the QPO frequencies with the propagation model. In this model, the harmonic is treated just like a QPO but with twice the frequency, however, for its physical interpretation it would be interesting to see the effect of second order variations of the parameters to be able to describe these components. Here, for simplicity we do not include reflection components in the spectral fitting since the timing model does not include the effect of reprocessing. This will be an important enhancement to be done in the future, which may provide a more accurate picture. Indeed, the timing  model used here has to be developed further to incorporate more complicated spectral models, that will allow for better comparison with high quality data.

\section{Acknowledgement}
We thank the anonymous reviewers for their helpful comments, which improved the quality of this work. We are grateful to the \textit{LAXPC} and \textit{SXT} teams for providing the data and requisite software tools for the analysis. NH is also grateful to the Inter-University Centre for Astronomy and Astrophysics (IUCAA) for giving the opportunity to visit in order to work on this project. NH acknowledges the funding received under the scheme of INSPIRE fellowship, Department of Science and Technology (DST). AG, SS and RM acknowledges the financial support provided by Department of Space, Govt of India No.DS$\_$2B-13012(2)/2/2022-Sec.2. 

\section{Data Availability}
Data analysed in this work is publicly available on Indian Space Science Data Center (ISSDC) website (\url{https://astrobrowse.issdc.gov.in/astroarchive/archive/Home.jsp}). The softwares used for \textit{LAXPC} and \textit{SXT} data reduction are available on \url{http://astrosat-ssc.iucaa.in/?q=laxpcData} and \url{http://www.tifr.res.in/~astrosat$\_$sxt/dataanalysis.html}, respectively. 

\bibliography{main} 

\end{document}